\documentclass[12pt]{iopart}
\usepackage{amsthm}
\usepackage{amsfonts}
\usepackage{amssymb}
\usepackage{bm}

\usepackage{mathptmx,times}

\newtheorem*{prop}{Proposition}

\newcommand{\dom}{\mathop{\rm dom}}
\newcommand{\sign}{\mathop{\rm sign}}
\newcommand{\spec}{\mathop{\rm spec}}

\newcommand{\Hat}{\hat}
\newcommand{\eqref}[1]{\textup{(\ref{#1})}}

\newcommand{\rr}{\mathbf{r}}

\newcommand{\mR}{\mathbb{R}}
\newcommand{\mC}{\mathbb{C}}
\newcommand{\mN}{\mathbb{N}}

\newcommand{\rR}{\mathrm{R}}
\newcommand{\rD}{\mathrm{D}}
\newcommand{\rJ}{\mathrm{J}}

\newcommand{\bA}{\mathbf{A}}
\newcommand{\ba}{\mathbf{a}}
\newcommand{\bK}{\mathbf{K}}
\newcommand{\bk}{\mathbf{k}}
\newcommand{\bp}{\mathbf{p}}

\newcommand{\cH}{\mathcal{H}}

\begin{document}

\title{Explicit Green functions for spin-orbit Hamiltonians}

\author{Jochen Br\"uning${}^1${}, Vladimir Geyler${}^{1}$\footnote{Deceased (2 April 2007)}  and Konstantin Pankrashkin${}^{1,2}$}

\address{${}^1$ Institut f\"ur Mathematik, Humboldt-Universit\"at zu Berlin, Rudower Chaussee 25, 12489 Berlin, Germany\\
${}^2$ L.A.G.A., Universit\'e Paris Nord, 99 av. J.-B. Cl\'ement, 93430 Villetaneuse, France (corresponding author)}
\ead{const@mathematik.hu-berlin.de}



\begin{abstract}
We derive explicit expressions for Green functions
and some related characteristics
of the Rashba and Dresselhaus Hamiltonians
with a uniform magnetic field.
\end{abstract}



\section{Introduction}
The Green function of a quantum Hamiltonian (integral kernel of the resolvent)
is one of the characteristics
whose knowledge usually permits to perform the complete spectral analysis and to study
various perturbations. The aim of the present communication is to obtain
explicit expressions for the Green function for a class of spin-orbit Hamiltonians,
namely, for the Rashba and Dresselhaus Hamiltonians with uniform magnetic fields,
whose study plays a central role in the spintronics~\cite{win}.
We use an abstract version of the construction~\cite{cs}, which permits
to reduce the problem to the well-known Green functions of the Landau Hamiltonian
and the Laplacian.

\section{Spin-orbit Hamiltonians}
Below we use the Pauli matrices
$\sigma_x=\left(
\begin{array}{cc} 0&1\\
                1&0\\
\end{array}
\right)$, $\sigma_y=\left(
\begin{array}{cc} 0&-\rmi\\
                \rmi&0\\
\end{array}
\right)$, $\sigma_z=\left(
\begin{array}{cc} 1&0\\
                0&-1\\
\end{array}
\right)$
and denote the identity $2\times2$-matrix by $\sigma_0$.

We consider Hamiltonians of a charged two-dimensional particle
in a uniform magnetic field $B$ orthogonal to the plane
and take into account the spin-orbit interaction.
Let $\bA$ be the magnetic vector potential, i.e.
$B=\displaystyle\frac{\partial A_y}{\partial x}-\displaystyle\displaystyle\frac{\partial A_x}{\partial y}$.
In what follows we use the symmetric gauge,
$A(x,y)=(\displaystyle\frac{By}{2},-\displaystyle\frac{Bx}{2})$.
Denote as usual $p_j:=-\rmi\hbar\nabla_j$ and $\Pi_j:=p_j-\displaystyle\frac{e}{c}\,A_j$,
$j=x,y$.
The Hamiltonian without spin-orbit interaction acts in the spinor space $L^2(\mR^2,\mC^2)$
and takes the form
$\Hat H_0=\displaystyle\frac{1}{2m_*}\bm{\Pi}^2\sigma_0$.
We are interested in the following two types of spin-orbit Hamiltonians.
The first one, the Rashba Hamiltonian $\Hat H_R$,
is of the form
\[
\Hat H_{\rR}=\Hat H_0+\displaystyle\frac{\alpha_{\rR}}{\hbar}\, \Hat U_{\rR}+\displaystyle\frac{g_*}{2}\,\mu_B B\sigma_z,\quad
\Hat U_{\rR}=\sigma_x\Pi_y-\sigma_y \Pi_x,
\]
where $\mu_B\equiv \displaystyle\frac{|e|\hbar}{2m_ec}$ is the Bohr magneton
($m_e$ is the electron mass), $g_*$ is the effective $g$-factor,
and $\alpha_{\rR}$ is the real-valued Rashba constant (whose dimension is $M L^3T^{-2}$)
expressing the strength of the spin-orbit interaction.
The second one, the Dresselhaus Hamiltonian, is given by
\[
\Hat H_{\rD}=\Hat H_0+\displaystyle\frac{\alpha_{\rD}}{\hbar}\, \Hat U_{\rD}+\displaystyle\frac{g_*}{2}\,\mu_B B\sigma_z,\quad
\Hat U_{\rD}=\sigma_y\Pi_y-\sigma_x \Pi_x,
\]
and $\alpha_{\rD}$ is the real-valued Dresselhaus constant (whose dimension is $M L^3T^{-2}$)
expressing the strength of the spin-orbit interaction.

In what follows we use mostly dimensionless coordinates introduced as follows.
Denote $\varkappa_{\rJ}:=\displaystyle\frac{m_*\alpha_{\rJ}}{\hbar^2}$, $\rJ=\rR,\rD$.
Furthermore, denote by $\Phi_0$ the magnetic flux quantum,
$\Phi_0:=\displaystyle\frac{2\pi \hbar c}{e}$, and let $b:=\displaystyle\frac{2\pi}{\Phi_0}B$ and
$\ba:=\displaystyle\frac{2\pi}{\Phi_0}\,\bA=(\displaystyle\frac{by}{2},-\displaystyle\frac{bx}{2})$. Now by setting $\bk:=\displaystyle\frac{1}{\hbar}\,\bp$,
$\bK:=\bk-\ba$ and introducing the coefficient $\gamma:=-\displaystyle\frac{g_*}{2}\,\displaystyle\frac{m_*}{m_e}$
we rewrite the above Hamiltonians as
$\Hat H_{\rJ}=\displaystyle\frac{\hbar^2}{2m_*}\,H_{\rJ}$, $\rJ=0,\rR,\rD$, where
\begin{eqnarray*}
H_0=\bK^2\,\sigma_0,\\
H_{\rR}=H_0+2\varkappa_{\rR} U_{\rR}+\gamma b \sigma_z, \quad U_{\rR}=\sigma_x K_y-\sigma_y K_x,\\
H_{\rD}=H_0+2\varkappa_{\rD} U_{\rD}+\gamma b \sigma_z, \quad U_{\rD}=\sigma_y K_y-\sigma_x K_x.
\end{eqnarray*}
In what follows we work with these new normalized Hamiltonians $H_{\rJ}$.

\section{Reduction to scalar case}

We start with a simple resolvent identity which is an abstract version
of the construction from~\cite{cs} and which is of crucial importance
in all our considerations.

For simplicity, for any self-adjoint operator $A$ and a complex number $E$
we use the notation $R(A,E):=(A-E)^{-1}$.
Now let $A$ be a self-adjoint operator acting in a certain Hilbert space,
$\alpha\in \mR$. Denote $B:=A^2+2\alpha A$.

Let $E\in\mC\setminus\spec B$, then
$(B-E)^{-1}=[(A+\alpha)^2-(E+\alpha^2)]^{-1}=
[(A+\alpha-\eta)(A+\alpha+\eta)]^{-1}$,
where
$\eta=\sqrt{E+\alpha^2}$ (here and below in this section $\sqrt{z}$ is a fixed continuous
branch of the square root on the complex plane $\mC$ with an
appropriate cut).

If both the numbers $\eta-\alpha$ and $-\eta-\alpha$ are outside of
$\spec A$ (in particular, if $\Im\,\eta\ne0$) and $\eta\ne0$, then
\[
\big((A+\alpha-\eta)(A+\alpha+\eta)\big)^{-1}= \displaystyle\frac{1}{2\eta}\big(
(A-\eta+\alpha)^{-1}-(A+\eta+\alpha)^{-1}\big).
\]
If, in addition, $-\eta+\alpha\notin \spec A$ and
$\eta+\alpha\notin \spec A$, then
$
(A-\eta+\alpha)^{-1}-(A+\eta+\alpha)^{-1}\nonumber\\
=(A+\eta-\alpha)(A^2-(\eta-\alpha)^2)^{-1}-
(A-\eta-\alpha)(A^2-(\eta+\alpha)^2)^{-1}$.
As a result, we arrive at the identity
\begin{equation}\fl
                     \label{2.5}
\begin{array}{rl}
R(B;E)&=\displaystyle
\frac{1}{2\eta}\big((A+\eta-\alpha)R(A^2;(\eta-\alpha)^2)-
(A-\eta-\alpha)R(A^2;(\eta+\alpha)^2) \big)\\[\medskipamount]
&= \displaystyle\frac{A-\alpha}{2\eta}\big(R(A^2;(\eta-\alpha)^2)-
R(A^2;(\eta+\alpha)^2)\big)\\[\medskipamount]
&\quad+ \displaystyle\frac{1}{2}\big(R(A^2;(\eta-\alpha)^2)+ R(A^2;(\eta+\alpha)^2)\big).
\end{array}
\end{equation}
Our aim now is to calculate the Green functions for $H_{\rR}$ and $H_{\rD}$
using Eq.~\eqref{2.5}.

Consider an operator $V_{\rJ}=U_{\rJ}+\beta_{\rJ}\sigma_z$, 
where $\beta_{\rJ}$ is a real constant which will be chosen later, $\rJ=\rR,\rD$.
Using the commutation relation $K_xK_y-K_yK_x=ib$ and the elementary properties
of the Pauli matrices one easily obtains
\begin{equation}
      \label{eq-VV}
V_{\rR}^2=(H_0+\beta_{\rR}^2)\sigma_0-b\sigma_z,
\quad
V_{\rD}^2=(H_0+\beta_{\rD}^2)\sigma_0+b\sigma_z.
\end{equation}
Therefore, 
\begin{equation}\fl
     \label{eq-V2H}
H_{\rJ}=V_{\rJ}^2+2\varkappa_{\rJ} V_{\rJ} -\beta_{\rJ}^2\sigma_0,\quad \rJ=\rR,\rD,
\quad \mbox{for} \quad
\beta_{\rR}:=\displaystyle\frac{\gamma+1}{2\varkappa_{\rR}}\,b,\quad \beta_{\rD}:=\displaystyle\frac{\gamma-1}{2\varkappa_{\rD}}\,b.
\end{equation}

\subsection{The free case}

We consider here the case without magnetic field, $b=0$. Then Eq.~\eqref{eq-V2H}
reads simply as $H_{\rJ}=U_{\rJ}^2+2\varkappa_{\rJ} U_{\rJ}$ with $U^2_{\rJ}=H_0\equiv -\Delta\,\sigma_0$.
Clearly, $\spec H_{\rJ}=f(\spec U_{\rJ})$ for $f(x)=x^2+2\varkappa_{\rJ}x$.
We see that $\spec U_{\rJ}=\mR$ (see appendix),
hence $\spec H_{\rJ}=[-\varkappa^2_{\rJ},+\infty)$ and the spectrum contains no eigenvalues.

Note that the Green function $G_0(x,y;z)$ of $H_0$ is known explicitly,
\[
G_0(\rr,\rr';z)=\displaystyle\frac{1}{2\pi} K_0\big(\sqrt{-z}|\rr-\rr'|\big),
\]
where $K_0$ is the McDonald function and $\sqrt{x}>0$ for $x>0$.
Hence, by \eqref{2.5}, the Green function $G_{\rJ}$, which is the integral
kernel of the resolvent $(H_{\rJ}-z)^{-1}$, has the form
\begin{eqnarray}\fl
G_{\rJ}(\rr,\rr';z)=\displaystyle\frac{1}{4\pi}\,\bigg[
\displaystyle\frac{1}{\rmi\sqrt{-(z+\varkappa_{\rJ}^2)}}(U_{\rJ}-\varkappa_{\rJ})
\big(
K_0(\zeta^+_{\rJ}|\rr-\rr'|)-K_0(\zeta^-_{\rJ}|\rr-\rr'|)
\big)\nonumber\\
+K_0(\zeta^+_{\rJ}|\rr-\rr'|)+K_0(\zeta^-_{\rJ}|\rr-\rr'|)
\bigg]\sigma_0,\quad \zeta^\pm_{\rJ}=\sqrt{-(z+\varkappa_{\rJ}^2)}\pm \rmi\varkappa_{\rJ},
             \label{eq-gr-nm}
\end{eqnarray}
where $U_{\rJ}$ acts as a differential expression with respect to $\rr$.
Applying the identity $K'_0(t)=-K_1(t)$ one concludes that
\[
G_{\rJ}(\rr,\rr';z)=\left(\begin{array}{cc}
G^{11}_{\rJ}(\rr,\rr';z) & G^{12}_{\rJ}(\rr,\rr';z)\\
G^{21}_{\rJ}(\rr,\rr';z) & G^{22}_{\rJ}(\rr,\rr';z)
\end{array}\right)
\]
with
\begin{eqnarray*}\fl
G^{11}_{\rJ}(\rr,\rr';z)=G^{22}_{\rJ}(\rr,\rr';z)= 
\displaystyle\frac{1}{4\pi}\,\bigg[
-\displaystyle\frac{\varkappa_{\rJ}}{\rmi\sqrt{-(z+\varkappa_{\rJ}^2)}}\times\\
\big( K_0(\zeta^+_{\rJ}|\rr-\rr'|)-K_0(\zeta^-_{\rJ}|\rr-\rr'|)
\big)
+K_0(\zeta^+_{\rJ}|\rr-\rr'|)+K_0(\zeta^-_{\rJ}|\rr-\rr'|)
\bigg]
\end{eqnarray*}
for both $\rJ=\rR,\rD$, and
\[\fl
G^{12}_{\rR}(\rr,\rr';z)=
\displaystyle\frac{\rmi(y-y')-(x-x')}{4\pi\, \rmi\sqrt{-(z+\varkappa_{\rR}^2)}\,|\rr-\rr'|}\times
\Big[
\zeta^+_{\rR}K_1\big(\zeta^+_{\rR}|\rr-\rr'|\big)
-
\zeta^-_{\rR}K_1\big(\zeta^-_{\rR}|\rr-\rr'|\big)
\Big],
\]
\[\fl
G^{12}_{\rD}(\rr,\rr';z)=
\displaystyle\frac{(y-y') -\rmi(x-x')}{4\pi\, \rmi\sqrt{-(z+\varkappa_{\rD}^2)}\,|\rr-\rr'|}
\times\Big[
\zeta^+_{\rD}K_1\big(\zeta^+_{\rD}|\rr-\rr'|\big)
-
\zeta^-_{\rD}K_1\big(\zeta^-_{\rD}|\rr-\rr'|\big)
\Big],
\]
and $G^{21}_{\rJ}(\rr,\rr';z)=\overline{G^{12}_{\rJ}(\rr',\rr;\bar z\,)}$.
We note that such a representation of the Green function was
essentially obtained in~\cite{cs} for special values of $z$.

\subsection{The magnetic case}
Using~\eqref{eq-V2H} one can easily calculate the spectrum of $H_{\rJ}$. Namely,
$\spec H_{\rJ}=g(\spec V_{\rJ})$ with $g(x)=x^2+2\varkappa_{\rJ}-\beta_{\rJ}^2$.
The spectrum of $V_{\rJ}$ can be calculated using the results of appendix.
Note that the spectrum of $H_0$ consists of the Landau
levels, $\spec H_0=\{|b|(2n+1):\,n\in\mN\}$.

Consider first the Rashba case. By~\eqref{eq-VV} we have
$\spec V_{\rR}^2=\{|b|(2n+1-s\sign b)+\beta_{\rR}^2:\,n\in\mN,\,\,s=\pm1\}$
and, respectively, $\spec\,V_{\rR}=\{\pm\sqrt{|b|(2n+1-s\sign b)+\beta_{\rR}^2}:\,n\in\mN,\,\,s=\pm1\}$.
Hence, the  spectrum of $H_{\rR}$ consists
of the Rashba levels, $\spec H_{\rR}=\{\varepsilon^{\pm}(n,s):\,n\in\mN,\,s=\pm1\}$,
$\varepsilon^{\pm}(n,s)=|b|(2n+1- s\sign b)\pm
2\varkappa_{\rR}\sqrt{\beta_{\rR}^2+|b|\big(2n+1- s\sign b\big)}$.

For the Dresselhaus case one has, exactly in the same way,
$\spec V_{\rD}^2=\{|b|(2n+1+s\sign b)+\beta_{\rD}^2:\,n\in\mN,\,s=\pm1\}$,
$\spec V_{\rD}=\{\pm\sqrt{|b|(2n+1+s\sign b)+\beta_{\rD}^2}:\,n\in\mN,\,\,s=\pm1\}$,
and the spectrum of $H_{\rD}$ consists
of the Dresselhaus levels, $\spec H_{\rD}=\{\varepsilon^{\pm}(n,s):\,n\in\mN,\,s=\pm1\}$,
$\varepsilon^{\pm}(n,s)=|b|(2n+1+s\sign b)\pm
2\varkappa_{\rD}\sqrt{\beta_{\rD}^2+|b|(2n+1+s\sign b)}$.
We note that the formulas for the eigenvalues were obtained e.g. in~\cite{zhang}
by a different method.

Now let us pass to the calculation of the Green functions.
Note that $H_0$ has the following Green function:
\begin{eqnarray*}\fl
G_0(\rr,\rr';z)=\frac{1}{4\pi}\,
\Gamma\big(\frac{1}{2}-\frac{z}{2|b|}\big)\times\\
\exp\Big(\frac{\rmi b}{2}(\rr\wedge\rr')-\frac{|b|}{4}(\rr-\rr')^2\Big)\,\Psi
\Big(\frac{1}{2}-\frac{z}{2|b|}\,,1;\frac{|b|}{2}(\rr-\rr')^2\Big),
\end{eqnarray*}
where $\Psi$ is the confluent hypergeometric function~\cite{erd}.
Clearly,
\begin{equation}\fl
                      \label{6.2}
(V^2_{\rR/\rD}-z)^{-1}=\left(\begin{array}{cc}
                                \big(H_0-(z-\beta_{\rR/\rD}^2\pm
                                b)\big)^{-1}&0\\
                                                             0&
                                \big(H_0-(z-\beta_{\rR/\rD}^2\mp b)\big)^{-1}
                                \end{array}\right),
\end{equation}
where $+/-$ corresponds to $\rR/\rD$.
Set now $\eta_{\rJ}:= \sqrt{z+\varkappa_{\rJ}^2+\beta_{\rJ}^2}$ and
\[
\zeta^{\pm}_{\rR}(b):=(\eta_{\rR}\pm\varkappa_{\rR})^2+ b -\beta^2_{\rR}\,,\quad
\zeta^{\pm}_{\rD}(b):=
(\eta_{\rD}\pm\varkappa_{\rD})^2- b -\beta_{\rD}^2.
\]
By \eqref{2.5}, we have
\begin{eqnarray*}\fl
(H_{\rJ}-z)^{-1}= \frac{V_{\rJ}-\varkappa_{\rJ}}{2\eta_{\rJ}}\,
\big((V_{\rJ}^2-(\eta_{\rJ}-\varkappa_{\rJ})^2)^{-1}-
(V_{\rJ}^2-(\eta_{\rJ}+\varkappa_{\rJ})^2)^{-1}\big)\\
+
\frac{1}{2}\,\big((V_{\rJ}^2-(\eta_{\rJ}-\varkappa_{\rJ})^2)^{-1}+
(V_{\rJ}^2-(\eta_{\rJ}+\varkappa_{\rJ})^2)^{-1}\big).
\end{eqnarray*}
Passing to the Green function, we obtain
\begin{eqnarray}\fl
G_{\rJ}(\rr,\rr';z)=\frac{U_{\rJ}+\beta_{\rJ}\sigma_z-\varkappa_{\rJ}}
{2\eta_{\rJ}}\times \nonumber\\
\fl\quad\left(\begin{array}{cc} G_0\big(\rr,\rr';\zeta_{\rJ}^-(b)\big)-
G_0\big(\rr,\rr';\zeta^+_{\rJ}(b)\big)&0\\
0&G_0\big(\rr,\rr';\zeta_{\rJ}^-(-b)\big)-
G_0\big(\rr,\rr';\zeta^+_{\rJ}(-b)\big)\end{array}\right)+ \nonumber\\
\fl \quad\frac{1}{2}\left(\begin{array}{cc}
G_0(\rr,\rr';\zeta_{\rJ}^-( b))+
G_0(\rr,\rr';\zeta^+_{\rJ}(b))&0\\
0&G_0(\rr,\rr';\zeta_{\rJ}^-(-b))+
G_0\big(\rr,\rr';\zeta^+_{\rJ}(-b)\big)
\end{array}\right). \label{eq-gr-mag}
\end{eqnarray}
Here $U_{\rJ}$ is considered as a differentiation
operator with respect to $\rr$. Using the identity $d\Psi(a,c,x)/dx=-a\Psi(a+1,c+1,x)$
one can write more explicit expressions for the Green function.
Namely,
\[
G_{\rJ}(\rr,\rr';z)=\left(\begin{array}{cc}
G^{11}_{\rJ}(\rr,\rr';z) & G^{12}_{\rJ}(\rr,\rr';z)\\
G^{21}_{\rJ}(\rr,\rr';z) & G^{22}_{\rJ}(\rr,\rr';z)
\end{array}\right)
\]
with
\begin{eqnarray*}\fl
G^{11}_{\rJ}(x,y;z)=
\frac{\beta_{\rJ}-\varkappa_{\rJ}}{2\eta_{\rJ}}
\Big(G_0\big(\rr,\rr';\zeta_{\rJ}^-(b)\big)- G_0\big(\rr,\rr';\zeta^+_{\rJ}(b)\big)\Big)+\\
 \frac{1}{2}\Big(
G_0\big(\rr,\rr';\zeta_{\rJ}^-( b)\big)+
G_0\big(\rr,\rr';\zeta^+_{\rJ}(b)\big)\Big),\\
\fl
G^{22}_{\rJ}(\rr,\rr';z)=
-\frac{\beta_{\rJ}+\varkappa_{\rJ}}{2\eta_{\rJ}}
\Big(G_0\big(\rr,\rr';\zeta_{\rJ}^-(-b)\big)- G_0\big(\rr,\rr';\zeta^+_{\rJ}(-b)\big)\Big)
+\\ \frac{1}{2}\Big(
G_0\big(\rr,\rr';\zeta_{\rJ}^-( -b)\big)+
G_0\big(\rr,\rr';\zeta^+_{\rJ}(-b)\big)\Big)
\end{eqnarray*}
for both $\rJ=\rR,\rD$,
\begin{eqnarray*}\fl
G^{12}_{\rR}(\rr,\rr';z)=|b|\,\Big((x-x')-\rmi(y-y')\Big)\,\bigg(\frac{\sign b -1}{2}\,
\Big[G_0\big(\rr,\rr';\zeta_{\rJ}^-(-b)\big)- G_0\big(\rr,\rr';\zeta^+_{\rJ}(-b)\big)\Big]\\
+\Big[F_0\big(\rr,\rr';\zeta_{\rJ}^-(-b)\big)- F_0\big(\rr,\rr';\zeta^+_{\rJ}(-b)\big)\Big]
\bigg),\\
\fl
G^{12}_{\rD}(\rr,\rr';z)=|b|\,\Big((y-y')-\rmi(x-x')\Big)\,\bigg(\frac{\sign b +1}{2}\,
\Big[G_0\big(\rr,\rr';\zeta_{\rJ}^-(-b)\big)- G_0\big(\rr,\rr';\zeta^+_{\rJ}(-b)\big)\Big]\\
-\Big[F_0\big(\rr,\rr';\zeta_{\rJ}^-(-b)\big)- F_0\big(\rr,\rr';\zeta^+_{\rJ}(-b)\big)\Big]
\bigg).
\end{eqnarray*}
and $G^{21}_{\rJ}(\rr,\rr';z)=\overline{G^{12}_{\rJ}(\rr',\rr;\bar z)}$, where
\begin{eqnarray*}
\fl F_0(\rr,\rr';z)=\displaystyle \frac{1}{4\pi}\,\Big(\frac{z}{2|b|}-\frac{1}{2}\Big)
\Gamma\big(\frac{1}{2}-\frac{z}{2|b|}\big)\times\\
\exp\Big(\frac{\rmi b}{2}(\rr\wedge\rr')-\frac{|b|}{4}(\rr-\rr')^2\Big)\,\Psi
\Big(\frac{3}{2}-\frac{z}{2|b|}\,,2;\frac{|b|}{2}(\rr-\rr')^2\Big).
\end{eqnarray*}

\section{Renormalized Green functions}

In some applications it is necessary to know the renormalized Green function, namely
the values $G^{\mathrm{ren}}_{\rJ}(\rr,\rr;z)$ given by
$G^{\mathrm{ren}}_\rJ(\rr,\rr;z)=\lim_{\rr'\to\rr}\Big[
G_\rJ(\rr,\rr';z)-S(\rr,\rr')
\Big]$,
where $S(\rr,\rr';z):=-\displaystyle\frac{1}{2\pi}\log|\rr-\rr'|\sigma_0$ is the on-diagonal singularity.
Terms of this kind appear e.g. when calculating the so-called Wigner $R$-matrix.

Consider first the case $b=0$. We will use the representation~\eqref{eq-gr-nm}
for the Green function. Using the expansion
\[
K_0(z)=\sum_{m=0}^\infty \frac{1}{(m!)^2}\,\Big(\frac{z^2}{2}\Big)^m
\big(
\psi(m+1)+\log 2-\log z
\big)
\]
one easily sees that $G^{12}_\rJ(\rr,\rr';z)$ are continuous functions
vanishing at $\rr=\rr'$, which means that $G^{\mathrm{ren}}_\rJ$ is diagonal.
To calculate the diagonal terms
we use the equality
\[
Q(z):=\lim_{r\to 0+}\frac{1}{2\pi}\Big(K_0(\sqrt{-z}\, r)+\log r\Big)=\frac{1}{2\pi}\big(\psi(1)-\frac12\log (-z)+\log 2\big).
\]
Hence, 
\begin{eqnarray*}
\fl
G^{\mathrm{ren}}_\rJ(\rr,\rr;z)=\bigg[-\frac{\varkappa_\rJ}{2 \rmi \sqrt{-(z+\varkappa_{\rJ}^2)}}\big(
Q(\zeta^+)-Q(\zeta^-)\big)+\frac{1}{2}\big(
Q(\zeta^+)+Q(\zeta^-)\big)\bigg]\sigma_0\\
=\frac{1}{2\pi}\bigg[
\psi(1)-\frac{1}{2}\log\big(-\frac{z}{4}\big)
+\frac{\varkappa_\rJ}{2\rmi \sqrt{-(z+\varkappa_{\rJ}^2)}}\,\log
\frac{\sqrt{-(z+\varkappa_{\rJ}^2)}+i\varkappa_\rJ}{\sqrt{-(z+\varkappa_{\rJ}^2)}-\rmi\varkappa_\rJ}
\bigg]\sigma_0,
\end{eqnarray*}
which is independent of $\rr$ due to the translational symmetry of the Hamiltonian.

For the magnetic case ($b\ne0$) we use the expansions~\cite{erd}
\begin{eqnarray*}
\fl
\Psi(a,n+1,x)=-\frac{(-1)^n}{\Gamma(a-n)}\,\Big[ \Phi(a,n+1,x)\log x\\
\fl \qquad
 + \sum_{r=0}^\infty \frac{(a)_r}{(n+1)_r r!} \big(
\psi(a+r)-\psi(1+r)-\psi(1+n+r)\big)x^r\Big] + \frac{(n-1)!}{\Gamma(a)}\sum_{r=0}^{n-1} \frac{(a-n)_r}{(1-n)_r} \frac{x\,^{r-n}}{r!},\\
\fl
\Phi(a,c,x)=\sum_{r=0}^\infty \frac{(a)_r}{(c)_r r!} x^r,\quad (a)_r:=\frac{\Gamma(a+r)}{\Gamma(a)}.
\end{eqnarray*}
The above expansions clearly show that the off-diagonal terms of $G^{\mathrm{ren}}_\rJ$
vanish. To express the diagonal terms we use the function 
\[\fl
Q(z):=\lim_{\rr\to \rr'}\Big(G_0(\rr,\rr';z)+\frac{1}{2\pi}\log |\rr-\rr'|\Big)=
-\frac{1}{4\pi}\Big(\psi\big(\frac{1}{2}-\frac{z}{2|b|}\big)-2\psi(1)+\log\frac{|b|}{2}\Big),
\]
then
\begin{eqnarray*}\fl
G_\rJ^{\mathrm{ren}}(\rr,\rr,z)=
\frac{\beta_{\rJ}\sigma_z-\varkappa_{\rJ}}
{2\eta_{\rJ}}\,
\left(\begin{array}{cc} Q\big(\zeta_{\rJ}^-(b)\big)-
Q\big(\zeta^+_{\rJ}(b)\big)&0\\
0&Q\big(\zeta_{\rJ}^-(-b)\big)-
Q\big(\zeta^+_{\rJ}(-b)\big)\end{array}\right)+\\
\frac{1}{2}\left(\begin{array}{cc}
Q(\zeta_{\rJ}^-( b))+
Q(\zeta^+_{\rJ}(b))&0\\
0&Q(\zeta_{\rJ}^-(-b))+
Q\big(\zeta^+_{\rJ}(-b)\big)
\end{array}\right).
\end{eqnarray*}

\appendix

\section*{Appendix. Supersymmetric spectral analysis}

\setcounter{section}{1}

For the sake of completeness, here we are going to prove the following
\begin{prop}\label{prop-susy}
Let $\cH_1$, $\cH_2$ be Hilbert spaces,
$A$ be a closed densely defined linear operator
from $\cH_1$ to $\cH_2$, and $m\ge 0$. On $\cH_1\oplus \cH_2$ consider the operator
$
L:=\left(\begin{array}{cc}
m & A^*\\
A & -m
\end{array}\right).
$
Then
\begin{equation}
        \label{eq-spL}
\spec L=-\sqrt{\spec(AA^*+m^2)}\cup \sqrt{\spec(A^*A+m^2)},
\end{equation}
and the same correspondence holds for the eigenvalues.
\end{prop}

\begin{proof}
First, it is well known~\cite{deift} that $\spec AA^*\setminus\{0\}=\spec A^*A\setminus\{0\}$.
Clearly, 
\begin{equation}
      \label{eq-LAA}
L^2=\left(\begin{array}{cc}
A^*A+m^2 & 0\\
0 & AA^*+m^2
\end{array}\right).
\end{equation}
Therefore, $\spec L^2\setminus\{m^2\}=\spec (AA^*+m^2)\setminus\{m^2\}$,
and for any $\lambda\in \spec AA^*\setminus\{0\}\equiv \spec AA^*\setminus\{0\}$ 
at least one of the numbers $-\sqrt{\lambda+m^2}$, $\sqrt{\lambda+m^2}$ lies in $\spec L$.
Let us show that actually they both are in the spectrum of $L$.

Let $\lambda>0$, $\lambda\in\spec A^*A$, then there exists a sequence $(\phi_n)$ with
$\phi_n\in\dom A^*A\subset\dom A$ such that
$\|\phi_n\|\ge 1$ and $\lim(A^*A-\lambda)\phi_n=0$. Denote
$
\psi_n:=\Big[\lambda+\big(\sqrt{\lambda+m^2}-m\big)\left(\begin{array}{cc}0 & A^*\\A & 0\end{array}\right)\Big]
\left(\begin{array}{c}
\phi_n\\0
\end{array}\right)$.
Clearly,
$\left(
\begin{array}{cc}
\phi_n\\0
\end{array}\right)
\perp \left(
\begin{array}{cc}
0 & A^*\\A & 0
\end{array}\right)
\left(\begin{array}{c}
\phi_n\\0
\end{array}\right)$,
which implies
\begin{equation}
     \label{eq-ll}
\|\psi_n\|\ge \lambda\|\phi_n\|\ge\lambda.
\end{equation}
By direct calculation, $
(L-\sqrt{\lambda+m^2}\,)\psi_n=\big(\sqrt{\lambda+m^2}-m\big)
\left(\begin{array}{c}
(A^*A-\lambda)\phi_n\\0
\end{array}\right)$.
Therefore, $\lim (L-\sqrt{\lambda+m^2}\,)\psi_n= 0$. Together with \eqref{eq-ll} this implies
$\sqrt{\lambda+m^2}\in\spec L$.

To show $-\sqrt{\lambda+m^2}\in\spec L$ one has to consider the functions
\[
\psi_n:=\bigg[\lambda-\big(\sqrt{\lambda+m^2}-m\big)\left(\begin{array}{cc}0 & A^*\\A & 0\end{array}\right)\bigg]
\left(\begin{array}{c}
0\\ \phi_n
\end{array}\right),
\]
where $\phi_n\in\dom AA^*\subset\dom A^*$, $\|\phi_n\|\ge 1$, and $\lim(AA^*-\lambda)\phi_n=0$ and to repeat the above steps. To finish the proof of Eq.~\eqref{eq-spL} it is necessary to study the points $\pm m$.
Not that it is sufficient to consider the situation when $\pm m$ is an isolated
point in $\spec L$ (in particular, an eigenvalue of $L$),
as for the continuous spectrum the result follows by
taking the closure of $\spec L\setminus\{-m,m\}$.

For $m=0$, Eq.~\eqref{eq-LAA}
reads as $\spec L^2=\spec AA^*\cup\spec A^*A$,
and the conditions $0\in\spec L$
and $0\in\spec AA^*\cup\spec A^*A$ are equivalent.

Assume $m\ne 0$ and $m$ is an eigenvalue of $L$, then there are
$\phi\in\dom A$ and $\varphi\in\dom A^*$ with $\|\phi\|+\|\varphi\|>1$ such that
$(L-m)\left(\begin{array}{c}
\phi\\
\varphi
          \end{array}\right)\equiv
\left(\begin{array}{c}
 A^*\varphi\\
A\phi -2m \varphi
\end{array}\right)=0$.
Clearly, this implies $A\phi_n\in\dom A^*$ and $A^*A\phi=0$.
If $\phi=0$, then also $\varphi=0$, which contradicts to the choice.
Therefore, $\phi$ is an eigenvector of $A^*A$.

Assume now that $0$ is an eigenvalue of $A^*A$, then there is a non-zero
vector $\phi\in\dom A^*A\subset\dom A$ with 
$\langle A^*A\phi_n,\phi_n\rangle\equiv \|A\phi\|=0$. Then
$(L-m)\left(\begin{array}{c}\phi\\0
\end{array}\right)=\left(\begin{array}{c}
0 \\
A\phi
\end{array}\right)=0$,
from which $m\in\spec L$.

The relationship between the conditions $-m\in L$
and $0\in\spec AA^*$  can be proved in a completely similar way.
\end{proof}

\ack This work was partially supported by the research fellowship of the Deutsche Forschungsgemeinschaft
(PA 1555/1-1) and SFB~647 (Berlin).

\end{document}